%% file: main.tex
\documentclass[10pt, conference, compsocconf]{IEEEtran}
\ifCLASSOPTIONcompsoc
\usepackage[caption=false,font=footnotesize,labelfont=sf,textfont=sf,skip=0pt]{subfig}
\else
\usepackage[caption=false,font=footnotesize]{subfig}
\fi
\usepackage[labelfont=bf,justification=centering,font=small,skip=2pt]{caption}
\usepackage{comment}
\usepackage[pdftex]{color, graphicx}
\usepackage{subfig}
\usepackage{multirow}
\usepackage{amsmath}
\usepackage{amssymb}
\usepackage[vlined]{algorithm2e}
\usepackage{url}
\usepackage{xcolor}
\usepackage{float}
\usepackage[T1]{fontenc}
\usepackage[latin1]{inputenc}
\usepackage{textcomp}
\usepackage{caption}		  

\usepackage[normalem]{ulem}
\usepackage{nicefrac}
\usepackage[normalem]{ulem}
\usepackage{soul}

\title{Understanding Cycling Mobility: Bologna Case Study}

\author{
\IEEEauthorblockN{Taron Davtian}
\IEEEauthorblockA{\textit{Institute of Computer Science} \\
\textit{University of Tartu}\\
Tartu, Estonia \\
taron.davtian@ut.ee}
\and
\IEEEauthorblockN{Flavio Bertini}
\IEEEauthorblockA{\textit{Department of Mathematical, Physical and} \\
\textit{Computer Sciences, University of Parma}\\
Parma, Italy\\
flavio.bertini@unipr.it}
\and
\IEEEauthorblockN{Rajesh Sharma}
\IEEEauthorblockA{\textit{Institute of Computer Science} \\
\textit{University of Tartu}\\
Tartu, Estonia \\
rajesh.sharma@ut.ee}
}

\begin{document}

\maketitle

\begin{abstract}
Understanding human mobility in urban environments is of the utmost importance to manage traffic and for deploying new resources and services. In recent years, the problem is exacerbated due to rapid urbanization and climate changes. In an urban context, human mobility has many facets, and cycling represents one of the most eco-friendly and efficient/effective ways to move in touristic and historical cities. The main objective of this work is to study the cycling mobility within the city of Bologna, Italy. We used six months dataset that consists of 320,118 self-reported  bike trips. In particular, we performed several descriptive analysis to understand spatial and temporal patterns of bike users for understanding popular roads, and most favorite points within the city. This analysis involved several other public datasets in order to explore variables that can possibly affect the cycling activity, such as weather, pollution, and events. The main results of this study indicate that bike usage is more correlated to temperature, and precipitation and has no correlation to wind speed and pollution. In addition, we also exploited various machine learning and deep learning approaches for predicting short-term trips in the near future (that is for the following 30, and 60 minutes), that could help local governmental agencies for urban planning. Our best model achieved an R square of 0.91, a Mean Absolute Error of 5.38 and a Root Mean Squared Error of 8.12 for the 30-minutes time interval.\newline\newline
\textit{keywords: }Cycling mobility, Geo-referenced data analysis, Big data, Short-term mobility forecasting, Machine Learning.
\end{abstract}

\IEEEpeerreviewmaketitle
\section{Introduction}\label{sec:intro}
\input{01.introduction}

\section{Related works}\label{sec:background} 
\input{02.related}

\section{Datsets}\label{sec:dataset}
\input{03.dataset}

\section{Descriptive Analysis}\label{sec:descrptive}
\input{04.descriptive}

\section{Predictive Analysis}\label{sec:predictive}
\input{05.predictions}

\section{Conclusions}\label{sec:conclusion}
\input{06.conclusion}

\section*{Acknowledgment}
We are grateful to SRM Reti e Mobilit\`a Srl for providing the data of the Bella Mossa program 2017. This research is financially supported by H2020 SoBigData++ and CHIST-ERA SAI project.

\bibliographystyle{IEEEtran}
\bibliography{07.bibliography.bib}
\label{sec:References}
\end{document}

%% file: 01.introduction.tex
In recent years, the possibility of geo-referenced data collection has opened doors for understanding human behaviour in urban contexts, such as pedestrian mobility \cite{mizzi2018unraveling}, taxi services \cite{rodrigues2018combining} and bike and scooter-sharing mobility \cite{zhu2020understanding}. In particular, bike-sharing services offer new opportunities for researchers to study human mobility by analyzing spatial patterns \cite{mooney2019freedom} and temporal patterns \cite{liu2018temporal} of bike usage or by studying the effects of weather on bike-sharing services \cite{nosal2014effect}.

Local governments promote cycling as an eco-friendly and healthy means of local transportation due to the significantly growing awareness about climate change and its capability in lightening road traffic within touristic and historical cities. However, to identify the transport demand of cyclists and to quantify the utilization of the road represent the main challenges to improve infrastructures (e.g., bike lanes and racks) and manage traffic.

Like many other cities, Bologna -a historical and a university city in Italy- struggles to manage urban traffic and its side-effect, such as CO2 emissions. In 2017, the Public Transport Authority \emph{SRM Reti e Mobilit\`a Srl}\footnote{http://www.srmbologna.it} promoted the six months \emph{Bella Mossa}\footnote{https://www.bellamossa.it} initiative through which it encouraged eco-friendly means of transportation and reduced day-to-day reliance on single-occupancy car journeys. The program was intended to reward users with points (to be transformed into prizes later) for walking, cycling, and using public transport. In particular, through a mobile application, the users chose the type of activity they were about to perform, and the application started tracking and recording positions via the users' smartphone GPS. In this study, we focused on cycling mobility due to the wider coverage of the city offered by this means of transportation (e.g., short/long trips, small/large streets).

To understand cycling mobility, we analyzed bike usage through the 320,118 self-reported unique bicycle trips in the city of Bologna within the \emph{Bella Mossa} dataset. These trips span across six months period, starting from April 2017 to September 2017. In particular, the study covers the following aspects:
\begin{enumerate}
   \item \textbf{Temporal analysis:} We evaluated how the cycling mobility changes during days, weeks and months, and computed the average speed, length, and travel times of the trips to understand the temporal characteristics of the cycling mobility.
    
    \item \textbf{Spatial analysis:} We reconstructed the utilization map of the city's road network highlighting popular roads and most favorite points within the city. The identification of the points of interest and routes used by users made it possible to study how people spread out from the main identified locations.

    \item \textbf{External variables:} To the best of our knowledge, when analyzing bike usage the studies in literature focused on a single aspect such as by either understanding the impact of weather conditions \cite{caulfield2017examining} or analysing temporal patterns \cite{liu2018temporal}. However, we considered several variables that can affect cycling mobility, such as weather, pollution, seasonality, holidays, and events.

    \item \textbf{Predictive analysis: } We used machine learning and deep learning approaches to forecast the number of trips for the near future, that is, in the next \emph{i)} 30 minutes, and \emph{ii)} 60 minutes to study daily bicycle traffic trends that could possibly help in traffic management and infrastructures deployment. 
\end{enumerate}

The paper is structured as follows. Section \ref{sec:background} outlines the current state of knowledge within which this study falls. Section \ref{sec:dataset} presents a detailed description of the datasets. In Section \ref{sec:descrptive}, we discuss the descriptive analysis results using the cycling dataset and other datasets about weather, pollution, and events. Section \ref{sec:predictive} presents the methodology and experimental results on short-term mobility forecasting. Finally, some concluding remarks and future works are made in  Section \ref{sec:conclusion}.

%% file: 02.related.tex
In this section, we examine works that have studied bike mobility. Firstly we discuss papers that have investigated the behaviors of the bike users (Section \ref{sec:rw:ana}). Then, we examine the papers where authors have performed predict analysis for different tasks, such as for distributing bikes in the city for easier usage or traffic flow (Section \ref{sec:rw:pred}).

\subsection{Analyzing bike usage}\label{sec:rw:ana}
Numerous studies have been done in order to understand the effect of weather on the use of bicycles. For example, in \cite{caulfield2017examining}, the research was conducted on how weather conditions affect the usage of bikes in the city of Cork, Ireland: trips are shorter during rains, and longer trips are done during sunny days. In \cite{nosal2014effect}, the authors found that precipitation in a single hour might significantly affect the number of bike trips. They also noted that cycling during weekends is more affected by weather than during weekdays. El-Assi et al. \cite{el2017effects} studied how weather affects bike sharing demand in the second biggest city of Canada, Toronto. The findings implied that there is a significant correlation between air temperature and bike usage. Authors also investigated how the built environment affects bike demand, concluding that bike infrastructure plays a major role in increasing the popularity of bike usage. Nankervis \cite{nankervis1999effect} found out in their study that weather's both short-term (e.g. daily temperature) and long-term (e.g. seasonal conditions) affect bike usage significantly. 
Similarly, in \cite{sears2012bike}, researchers discussed how seasonal factors affect bicycle commuting. The results of the study confirmed that there is a high correlation between weather and bike usage.

Some other studies focus on the impact of infrastructure on bike usage. For example, in \cite{liu2018temporal}, the authors investigated which road attributes influence bike users' path choices. In \cite{fishman2015factors}, researchers in Melbourne and Brisbane tried to quantify the factors influencing bike-share membership through a questionnaire. The results showed that the distance to the closest docking station is highly correlated with the membership and the authors indicated that the result confirms other prior researches \cite{fuller2011use}, \cite{molina2015bicycling}.

Some other works analysed bike usage across different parts of the cities. For example, in \cite{froehlich2009sensing} Froehlich et. al. investigated behaviors of bike users across different locations and times of the day in Barcelona's neighborhood. In \cite{zhao2015exploring}, the authors explored the bike-sharing travel time and trips by gender and day of the week. The results of the study suggested that demand for bike usage is generated in the residential districts, while the biggest hubs are train stations. The paper showed that there is a big difference in the sense of distance and trip duration between men and women. Additionally, the authors reported that women are using bikes during weekdays more, while men use it on average more during weekends. In another paper, researchers examined the behavior of bike usage in the city of Lyon, France, by focusing on how social behavior help in planning and designing policies in transportation \cite{borgnat2011shared}.

In this work, we analysed how different indicators such as i) air pollution, ii) weather conditions, iii) temperature affect bike usage. In addition, we also analysed the spatial and temporal dynamics of bicycle usage in order to provide a holistic picture of cycling mobility.

\subsection{Predictions using bike data}\label{sec:rw:pred}
The second line of our research has been conducted about predicting the usage of bikes by exploring various machine learning algorithms. In \cite{xu2013public}, authors analyzed operational data from bike-sharing systems to understand activity patterns and to use these patterns in order to plan the distribution of bikes. A hierarchical prediction model was employed in \cite{li2015traffic} with an aim to predict the number of bikes that are rented from each station to meet the demand. In contrast, Xu et al. \cite{xu2013public} used a hybrid prediction model by combining clustering with a support vector machine for predicting bicycle traffic flow. In a different work \cite{zhang2016bicycle}, the authors tried to predict the destination and arrival time of each bicycle trip, which can effectively help the companies to move bikes on time for the under-supplied station. 

In some recent works, a deep learning approach has been used in order to forecast bike demand. Some of them experimented with large-scale datasets and tried to predict demand for different time intervals using only historical bike usage data \cite{ai2019deep}, \cite{xu2018station}. Zhang et al. \cite{zhang2018short} used Long Short-Term Memory (LSTM) model to predict the number of trips by also considering public transport usage. For predicting the facility choice of cyclists between on-street and off-street facilities, a machine learning model was developed by authors of \cite{duc2018modeling}. In another work, along with the usage of weather data, taxi usage was also explored for predicting bike trips for the city of New York \cite{singhvi2015predicting}.

Recently, a number of studies have also investigated dockless bike-sharing systems \cite{luo2019comparative}, \cite{mckenzie:LIPIcs:2018:9374}, \cite{shaheen2019shared}. The main emphasis is on analysing spatial patterns of such systems, as without stations, bicycles can be left anywhere in the city, which also raises the question of redistribution of bikes in particular during the weekdays \cite{mooney2019freedom}, \cite{gu2019or}, \cite{shen2018understanding}. These papers have focused mainly on the problem of redistribution of the bikes in different locations. In addition, they tried to predict the locations where there will be a surplus of bikes as well as locations having a shortage of bikes. In this work, we tried to predict the number of trips by not only considering the historical bike usage data but also air temperature, precipitation, holidays, and events.

%% file: 03.dataset.tex
This section of the paper provides information about the datasets we used for the analysis. In particular, we describe the \emph{Bella Mossa} 2017 dataset about self-reported bike usage, and the other supplementary data sources (weather, pollution, events, etc.) that were used for our analysis.

\subsection{Bella Mossa 2017 dataset}
The main dataset used in the paper contains mobility data of different transportation means for the period of six months, that is, from April 1, 2017 to September 30, 2017. \emph{Bella Mossa} was an initiative promoting a healthy lifestyle and sustainable mobility among users residing in Bologna, Italy. It gave the user a chance to win various gifts and discounts as a reward for using a more sustainable and healthy means of transportation. For participating, a user just needed to download the mobile application and started it whenever they go out for a walk, use bikes, trains, buses, or even when using carpooling. The running application tracked and stored data related to the users' position using the GPS of the users' smartphone.
During the six months of the program, there were over 15,000 unique users who registered for this program, and 3.7 million Km was covered by them during 895,000 journeys. For security reasons, the dataset does not include users' personal information, and thus, we could not distinguish and analyze different patterns considering users gender or age. Moreover, the data is anonymised, and it is worth noting that each trip in the dataset is identified by a unique identification number (ID), however, this ID does not allow users to be identified on different days.

In the paper, we focused on the bike usage, that during the six months have generated 72,398,780 data points in 320,118 unique trips. Each record of the dataset is characterized by the following attributes: the activity ID (to identify the points of the same trip), the timestamp, the geographic coordinates (i.e., latitude and longitude), the GPS accuracy, and the speed. We grouped the data points according to the activity ID and the timestamp. The missing values were replaced by interpolation of the previous and next values during the trip. We used trips' information to analyze factors affecting the number of trips, while we used raw data points to reveal which streets and parts of the cities had been mostly utilized by bike users. In particular, we analyzed bike usage patterns during different periods, weather conditions and social events. We also incorporated these features for forecasting the number of trips for the short term prediction (that is, in the next 30 and 60-minutes time interval) for understanding how busy specific streets would be in terms of bike usage.

\subsection{Supplementary data sources}
Besides the main dataset, we used several supplementary datasets for analyzing bike usage and predicting the number of bike trips.

\noindent \textbf{1) Weather:} We downloaded historical data about temperature, precipitation, and wind for the observation period for the city of Bologna from the website of the Regional Agency for Prevention, Environment, and Energy\footnote{https://simc.arpae.it/dext3r} of Emilia-Romagna, Italy.  
The dataset contains information about the hourly average air temperature above 2 meters from the ground, the hourly average wind speed above 10 meters from the ground, and the cumulative precipitation data over 1 hour period.

\noindent \textbf{2) Pollution:} We used the information about pollution from the website of the Ministry of Economic Development\footnote{https://www.sviluppoeconomico.gov.it} in Italy. Based on the 2005 World Health Organization (WHO) Air Quality Guidelines\footnote{https://www.who.int/}, we analysed four main air pollution indicators, that is i) Particulate matter (PM), ii) Ozone (O3), iii) Nitrogen dioxide (NO2), and iv) Sulfur dioxide (SO2).

\noindent \textbf{3) Holidays and events:} We used publicly available information about the holidays in Italy and public events in the city of Bologna\footnote{http://presidenza.governo.it}. The dataset contains information about public holidays, national celebration days, and civil solemnities in the city of Bologna. Also, we gathered information about strikes and protests for the observation period for the city of Bologna\footnote{http://scioperi.mit.gov.it/mit2/public/scioperi/ricerca}.

%% file: 04.descriptive.tex
In this section, we present the results of the descriptive analysis of the \emph{Bella Mossa} bike data. Firstly, we present the results of the temporal analysis looking into daily, monthly and seasonal trends of bike usage (Section \ref{sec:tempAna}). Then, we show the spatial analysis results to highlight which parts of the city attract most of the bike trips (Section \ref{sec:SpaAna}). Next, we present the results by combining the cycling mobility data with the supplementary data sources. In particular, we discuss how weather conditions, such as a change in air temperature and precipitation, correlates with bike usage (Section \ref{sec:weaAna}). Also, we investigated the changes in three different indicators of air pollution with bike usage (Section \ref{sec:subsec_pollution}). Finally, we analysed public holidays and events, such as strikes and protest, in relation to bike usage (Section \ref{sec:eveAna}).

\begin{figure*}[t]
	\subfloat[]{
		\includegraphics[width=0.315\textwidth]{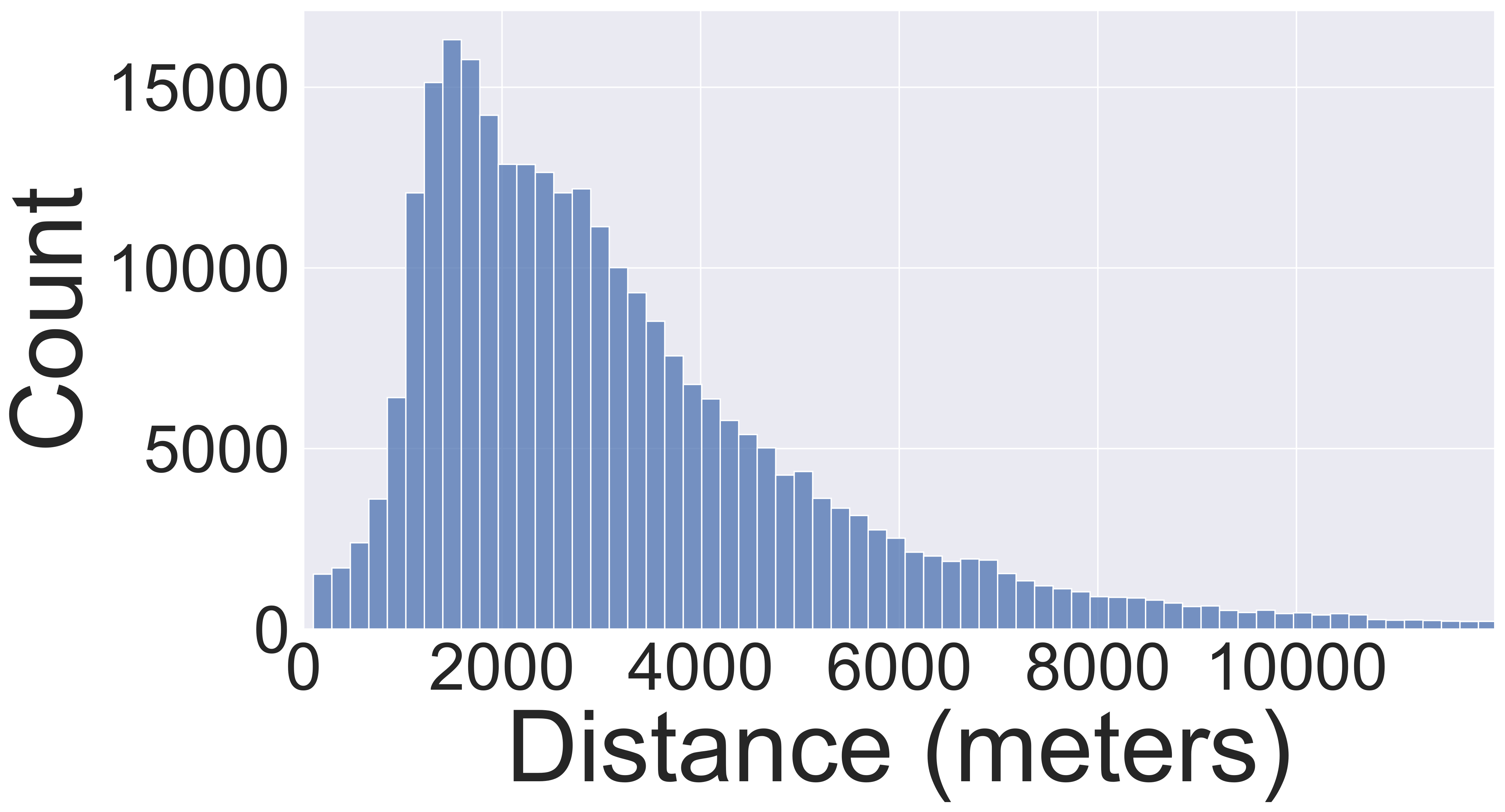}\label{fig:distanceTrend}
	}~\hfill
	\subfloat[]{
		\includegraphics[width=0.315\textwidth]{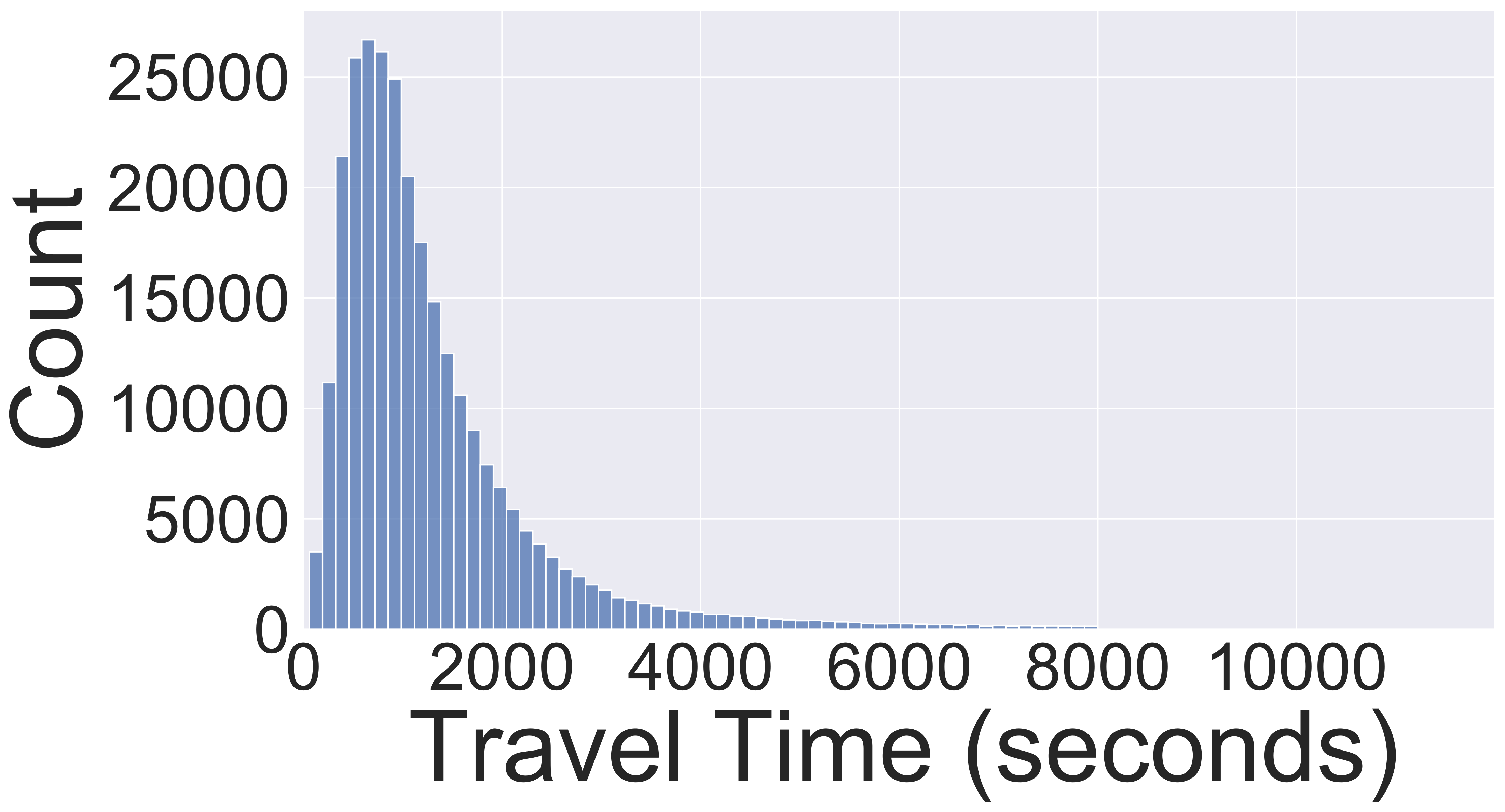}\label{fig:timeTrend}
	}~\hfill
	\subfloat[]{
		\includegraphics[width=0.315\textwidth]{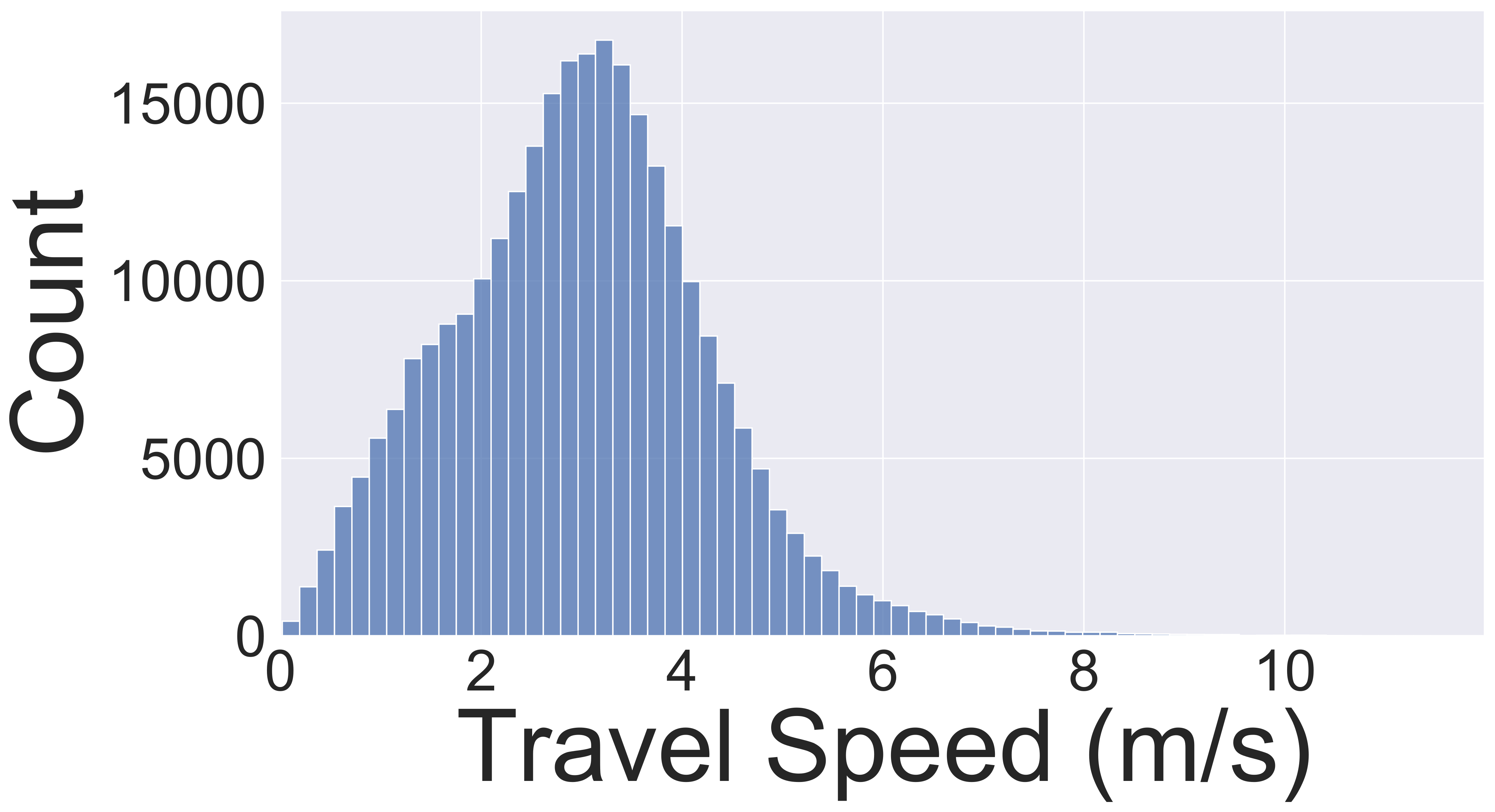}\label{fig:speedTrend}
	}
	\caption{Distribution of the covered distances (a), the travel times (b) and the average travel speeds (c).}
	\label{fig:accuracy_fast_bc}
\end{figure*}

\subsection{Temporal analysis}\label{sec:tempAna}
Firstly, for the whole observation period, we computed the i) distance distribution, ii) travel times distribution, and iii) average speeds of the trips. We also analysed both daily and weekly usage of bikes, which helped us in understanding long term trends, such as monthly and seasonal trends. In particular, the seasonality study allowed us to analyse the change in mobility of bicycle users during spring, summer, and the start of autumn.

Figures \ref{fig:distanceTrend} and \ref{fig:timeTrend} provide the distribution of the covered distances and travel times, showing a peak around 1600 meters and 720 seconds, respectively. This highlights a characteristic of the  bike usage in Bologna that 40\% of the trips cover less than 1.6 kilometers distance for 12-13 minutes of travel time. Whereas, Figure \ref{fig:speedTrend} shows the histogram of the average travel speeds with a peak of around 3.8-4 meters per second, that is, compatible with the speed usually reported in experimental literature \cite{dozza2014introducing}, \cite{menghini2010route}.

\begin{figure}[!t]
    \centering
    \includegraphics[width=1\columnwidth]{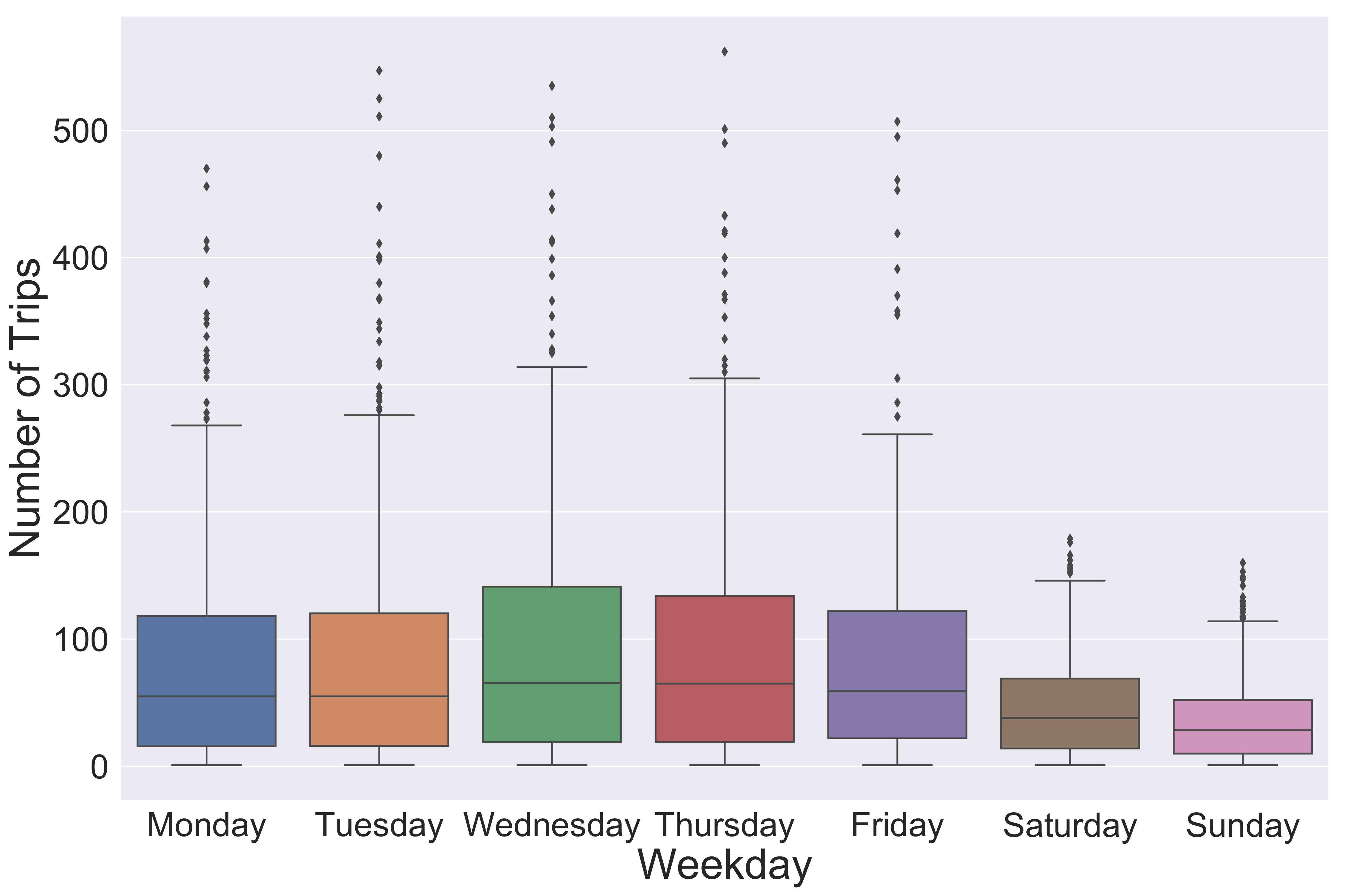}
    \caption{Number of trips aggregated by weekdays.}
    \label{fig:Weekly trend}
\end{figure}

\begin{figure}[!t]
    \centering
    \includegraphics[width=1\columnwidth]{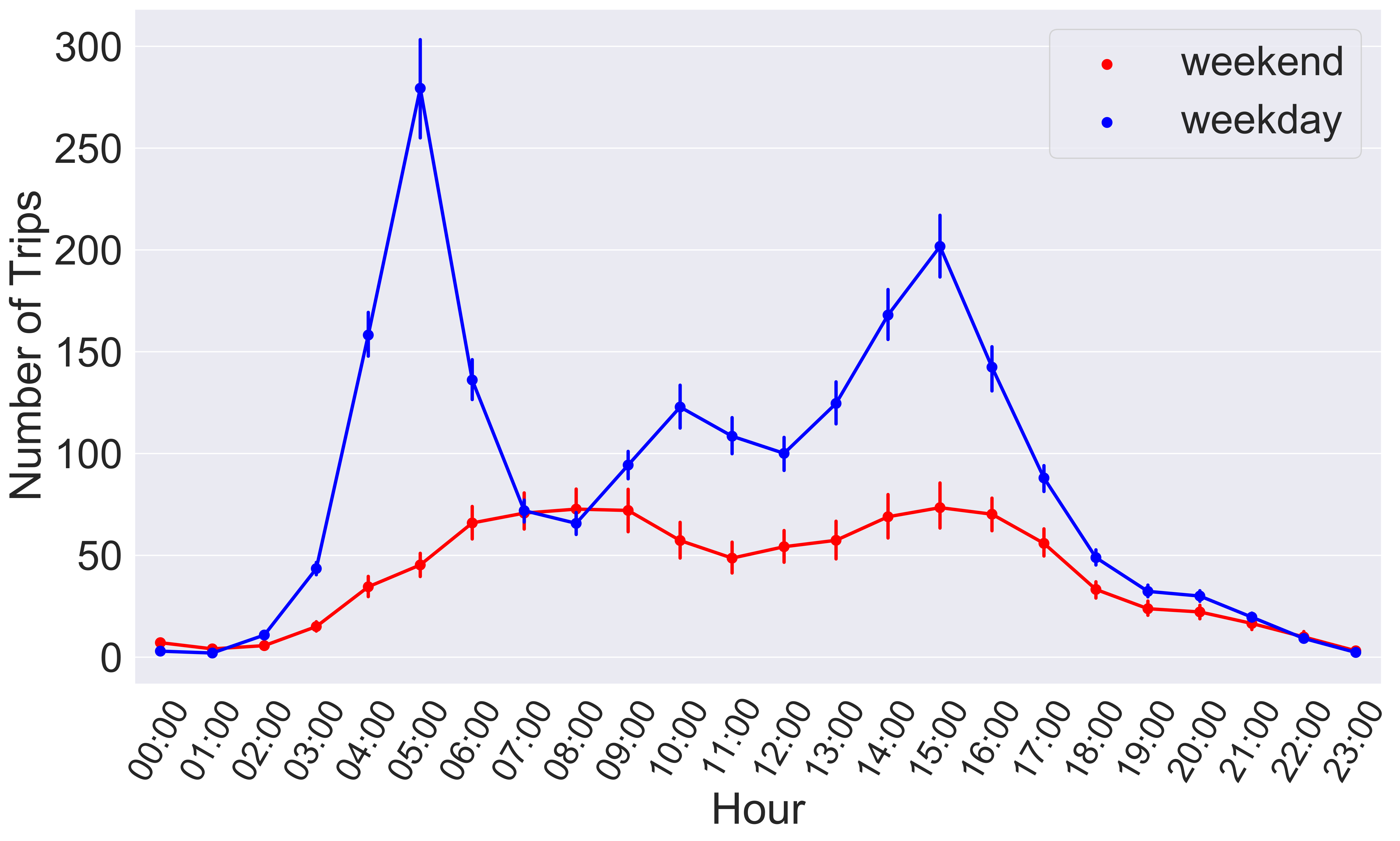}
    \caption{Weekends vs weekdays: the start and end times of work represent two peaks during the week.}
    \label{fig:Daily trend}
\end{figure}

\begin{figure}[!t]
    \centering
    \includegraphics[width=1\columnwidth]{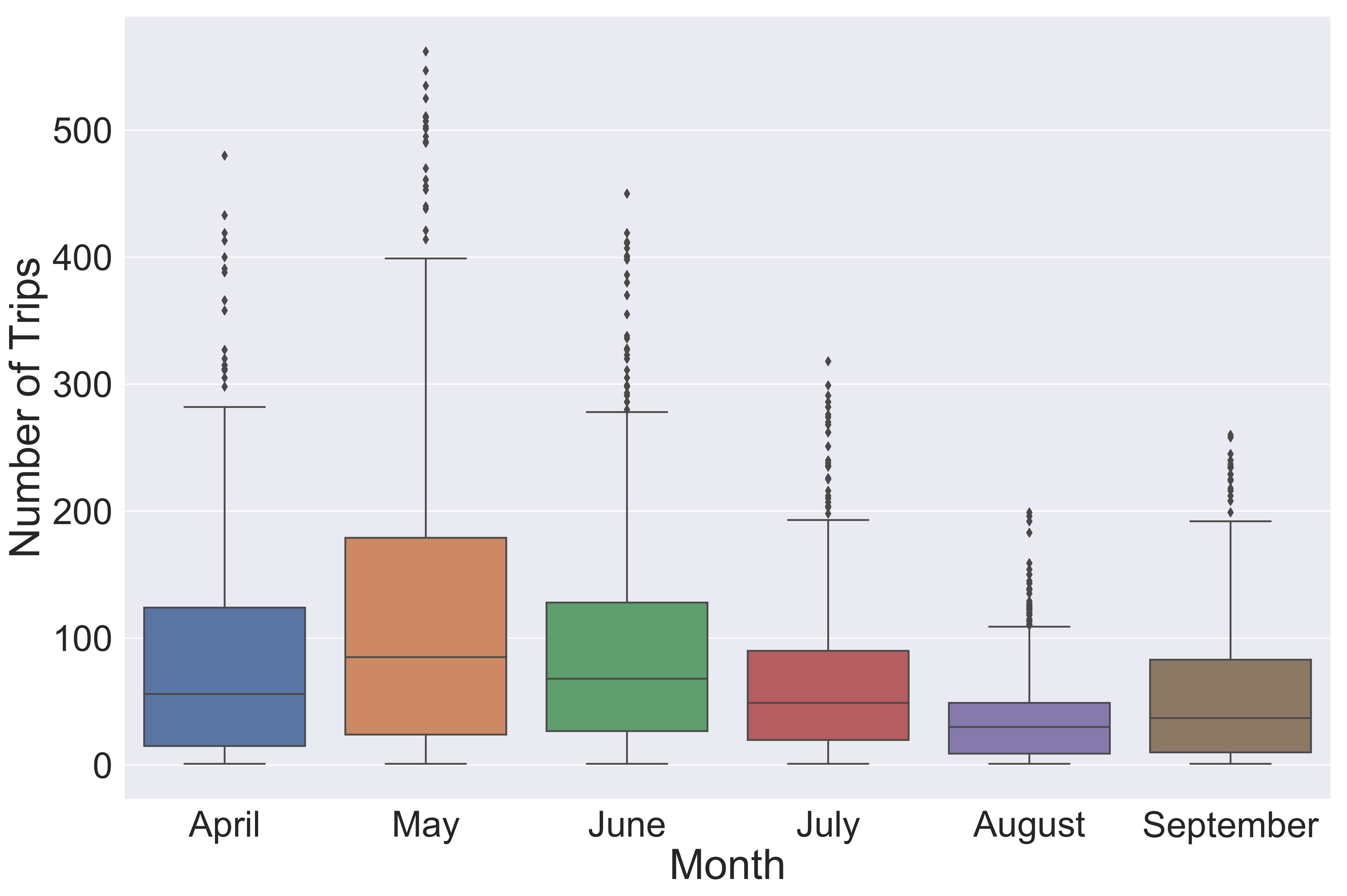}
    \caption{Seasonality of cycling mobility.}
    \label{fig:Monthly trend}
\end{figure}

Figure \ref{fig:Weekly trend} shows the trips aggregated by weekdays. The 84\% of total trips done during working days suggests that on weekdays bikes are heavily used, possibly either for home-to-work or home-to-study. This assumption is also confirmed in Figure \ref{fig:Daily trend} as we can see that the most of the trips during the working days were happening from 07:00 to 09:00 and from 17:00 to 19:00, that is, mostly connected with the time people move for either work or study and returning in the evening. 

Figure \ref{fig:Monthly trend} shows the number of trips spread over the six months. We can observe that the number of trips reached its peak during May (25\%) and then it gradually dropped down, reaching its lowest level during August. In particular, there was a 40\% drop in trips from July to August. In August, only 30\% of May's trips were done. This behaviour can be explained as during summer more, and more people start leaving the city for vacations, and the off-site students start going back to their hometowns. However, in September the number of trips started increasing again, with an increase of 28\% compared to the number of trips in August, possibly due to the return of students and people from holidays.

\subsection{Spatial analysis}\label{sec:SpaAna}

\begin{figure*}[ht!]
\centering
\begin{minipage}[b]{0.30\linewidth}
   \includegraphics[width=0.95\columnwidth]{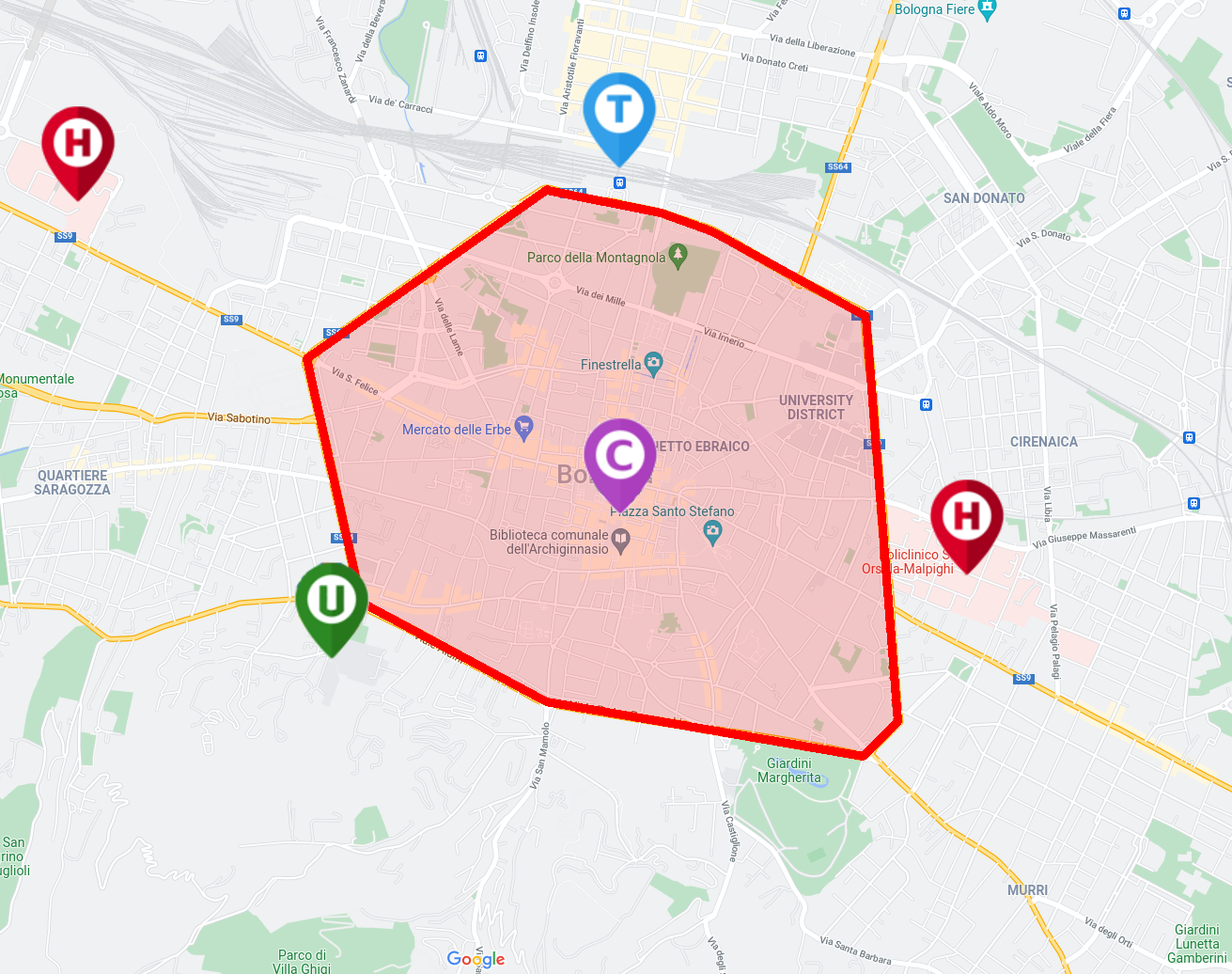}
    \caption{Bologna Map.}
    \label{fig:bologna}
\end{minipage}
\quad
\begin{minipage}[b]{0.30\linewidth}
  \includegraphics[width=0.95\columnwidth]{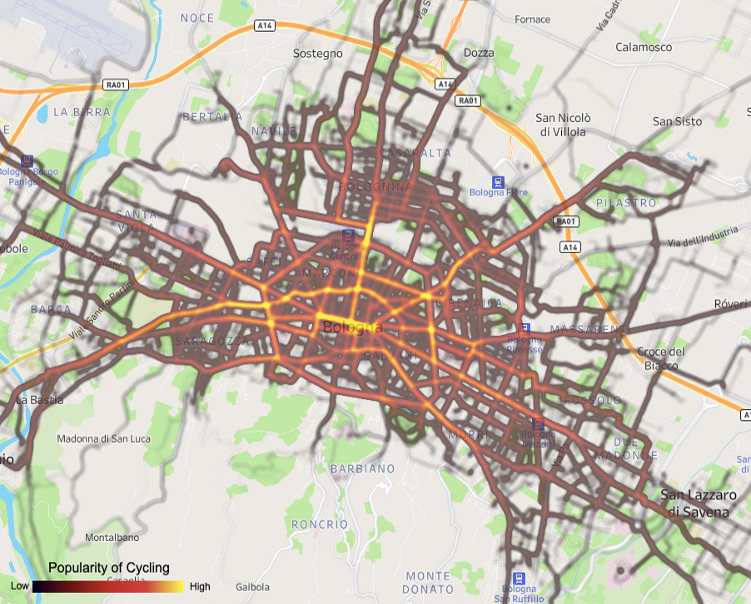}
    \caption{Bike road network in May}
    \label{fig:density_full_a}
\end{minipage}
\quad
\begin{minipage}[b]{0.30\linewidth}
 \includegraphics[width=0.95\columnwidth]{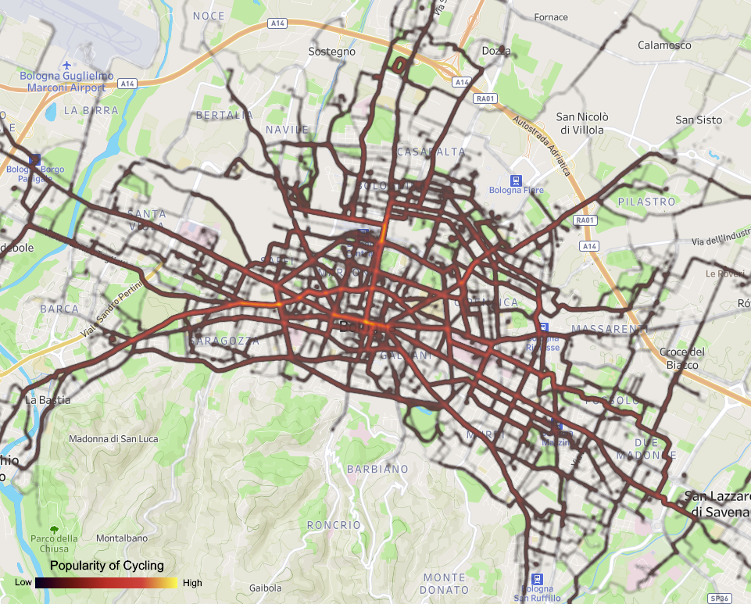}
    \caption{Bike road network in August}
    \label{fig:density_full_b}
\end{minipage}
\end{figure*}

In this section, we discuss the paths used by bike users and the possible attractive hubs of the city. Bologna's historical city centre covers an area of 4.5Km$^2$, which is the red polygon shown in Figure \ref{fig:bologna}. The city has several characteristics that can help to understand urban mobility. In particular, the train station is located north of the city (denoted using ``T'' letter in light blue color in Figure \ref{fig:bologna}) just outside the historic centre (denoted using ``C'' letter in purple color in Figure \ref{fig:bologna}), and it is one of the main nodes in the national railway network. Most of the departments of the University of Bologna are distributed within the historic centre, however, the engineering department is located on the south west part of the city (denoted using ``U'' letter in green color in Figure \ref{fig:bologna}). There are two big hospitals in the east and west of the city (denoted using ``H'' letter in red color in Figure \ref{fig:bologna}). During the day, there is a large number of people who move to and from the city for work and study purposes.

\subsubsection{\textbf{Bike road network}}
In this section, we identify the road networks often used by the bike users (based on the dataset). Figures \ref{fig:density_full_a} and \ref{fig:density_full_b} show which streets were the most used by bike users through the density-based gradient palette for the months of May 2017 and August 2017. For each month, we normalised the data before plotting the graphs. The yellow roads were the most used, while the red ones were used by a smaller number of users. We can clearly observe the difference in terms of bike usage between the two months. We can observe that the density of the streets is similar, and the high-used paths correspond to the main arteries of the centre's road network. The road network in Bologna has a radial structure and, since the bicycle is used for medium-long trips, it is quite common to cross the city passing through the centre.

\subsubsection{\textbf{Main hubs}} 
\begin{figure}[t]
	\subfloat[]{
		\includegraphics[width=0.45\textwidth]{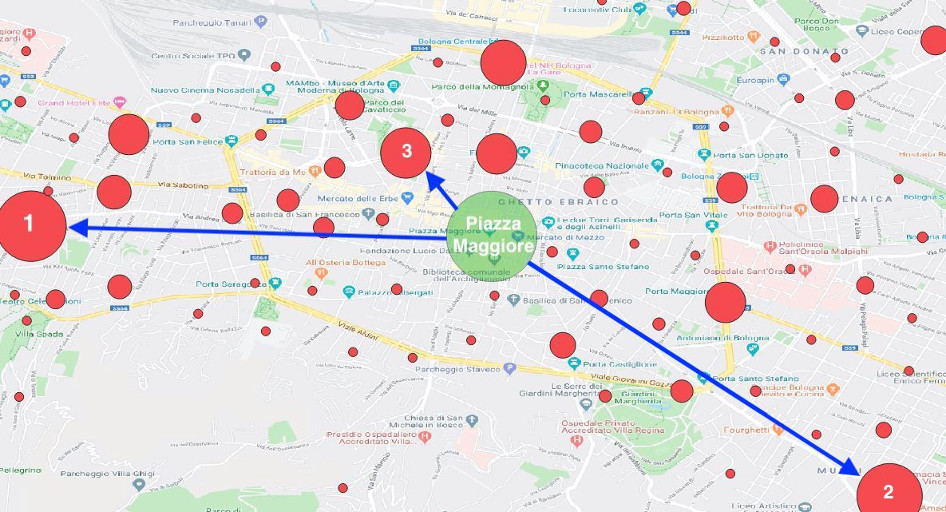} 
    \label{fig:centre_hub_a}
	}~\hfill \\
	\subfloat[]{
	\includegraphics[width=0.45\textwidth]{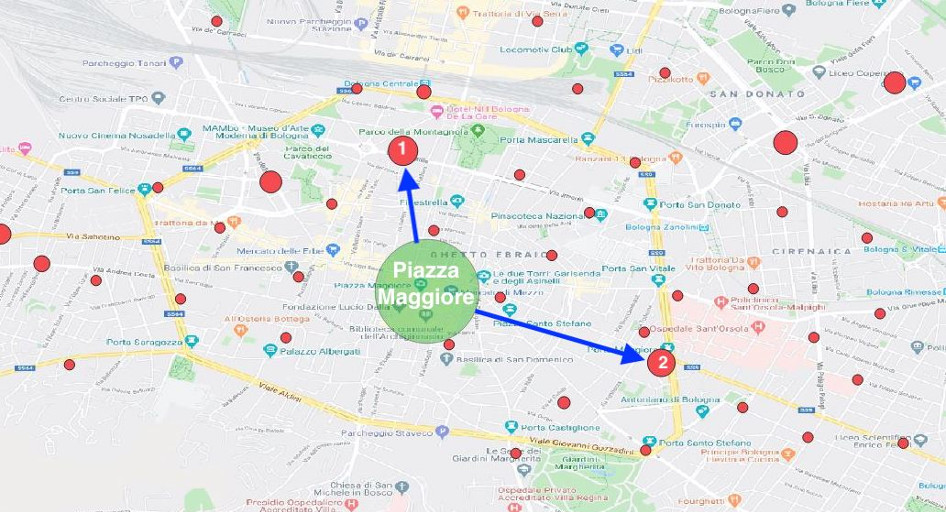}
       \label{fig:centre_hub_b}
	}
\caption{Spreading out pattern from the \emph{Piazza Maggiore} hub, difference between May (a) and August (b).}
  \label{fig:centre_hub}
\end{figure}

The analysis of the trips allowed us to identify three main hubs of the city, which have a high number of either start or end points. First being the \emph{Piazza Maggiore} -- the main square and the neighborhood streets, which are one of the main centers of city life. The other two are the train station, and the engineering department of the University of Bologna. To understand how bike trips spread out from the  three aforementioned hubs, we compared if there is a significant change in bike usage between the month of May (chosen as one of the representatives of the non-holiday period) and the month of August (holiday period) for each hub. In Figures \ref{fig:centre_hub}, \ref{fig:train_hub}, and \ref{fig:unibo_hub}, the locations were aggregated by the number of trips started from the main hub (the green circles in each figure) and ended in the final destination (the ranked red dots in the figures). The circle size indicates the number of trips ending at that destination. In addition, we also ranked these locations by inserting numbers in the circles.

\begin{enumerate}

\item \noindent The spreading pattern from \textbf{Piazza Maggiore} (in the city centre) changed considerably between May and August. May is considered a ``working month'' and we can see that the main destinations were outside the city centre (Figure \ref{fig:centre_hub_a}). This indicates that most probably people were going back to home after spending time in the city centre (possibly from the workplaces and offices in the city center). Indeed, most of the trips ended in residential areas of the city. In contrast, during August, which is a ``vacation month'' most of the trips ended inside historical city centre (Figure \ref{fig:centre_hub_b}).

\begin{figure}[t]
	\subfloat[]{
		\includegraphics[width=0.45\textwidth]{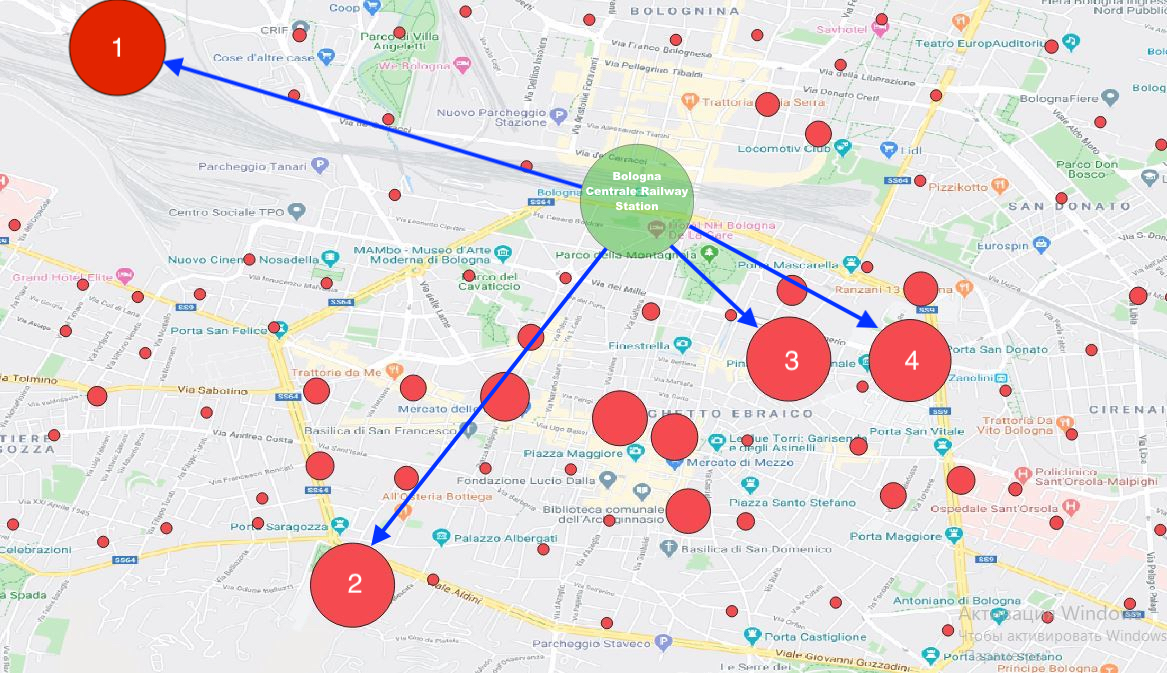}
		\label{fig:train_hub_a}
	}~\hfill \\
	\subfloat[]{
		\includegraphics[width=0.45\textwidth]{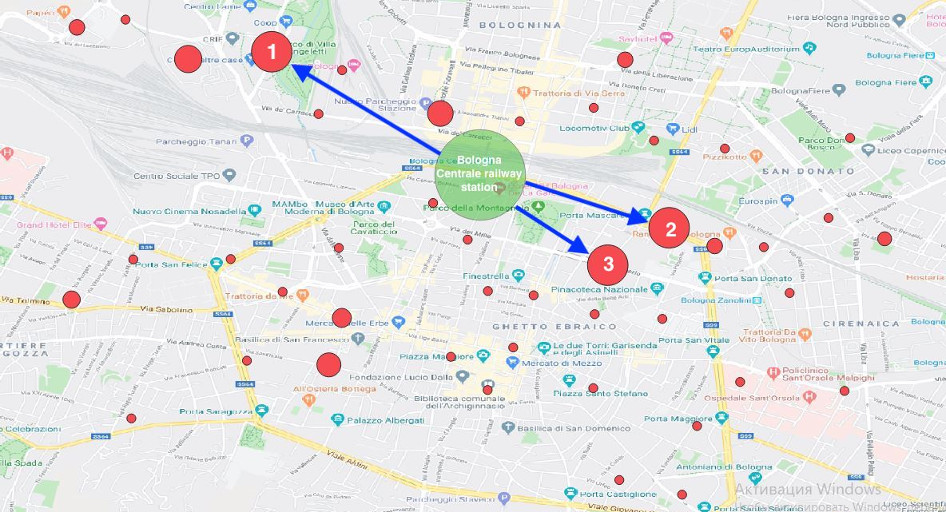}
		\label{fig:train_hub_b}
	}
 \caption{Spreading out pattern from the train station hub, difference between May (a) and August (b).}
  \label{fig:train_hub}
\end{figure}

\begin{figure}[t]
	\subfloat[]{
		\includegraphics[width=0.45\textwidth]{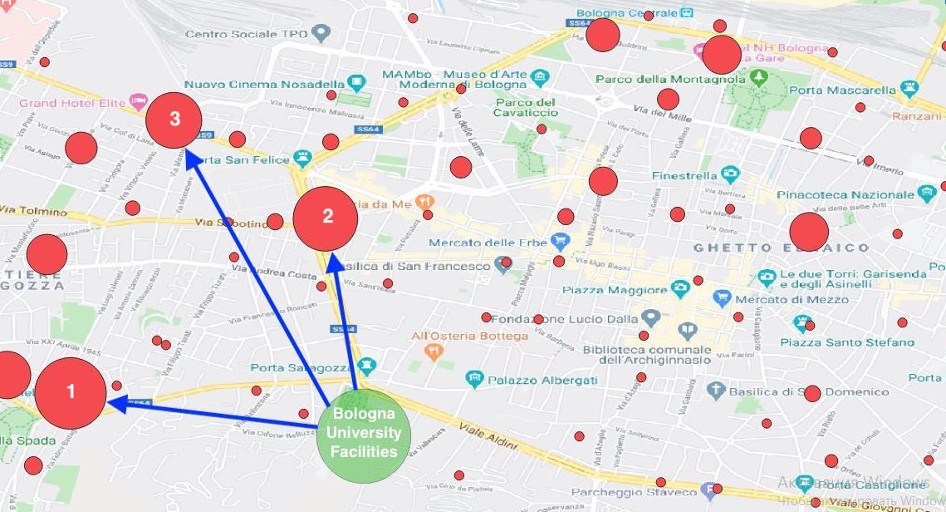}\label{fig:unibo_hub_a}
	}~\hfill \\
	\subfloat[]{
	\includegraphics[width=0.45\textwidth]{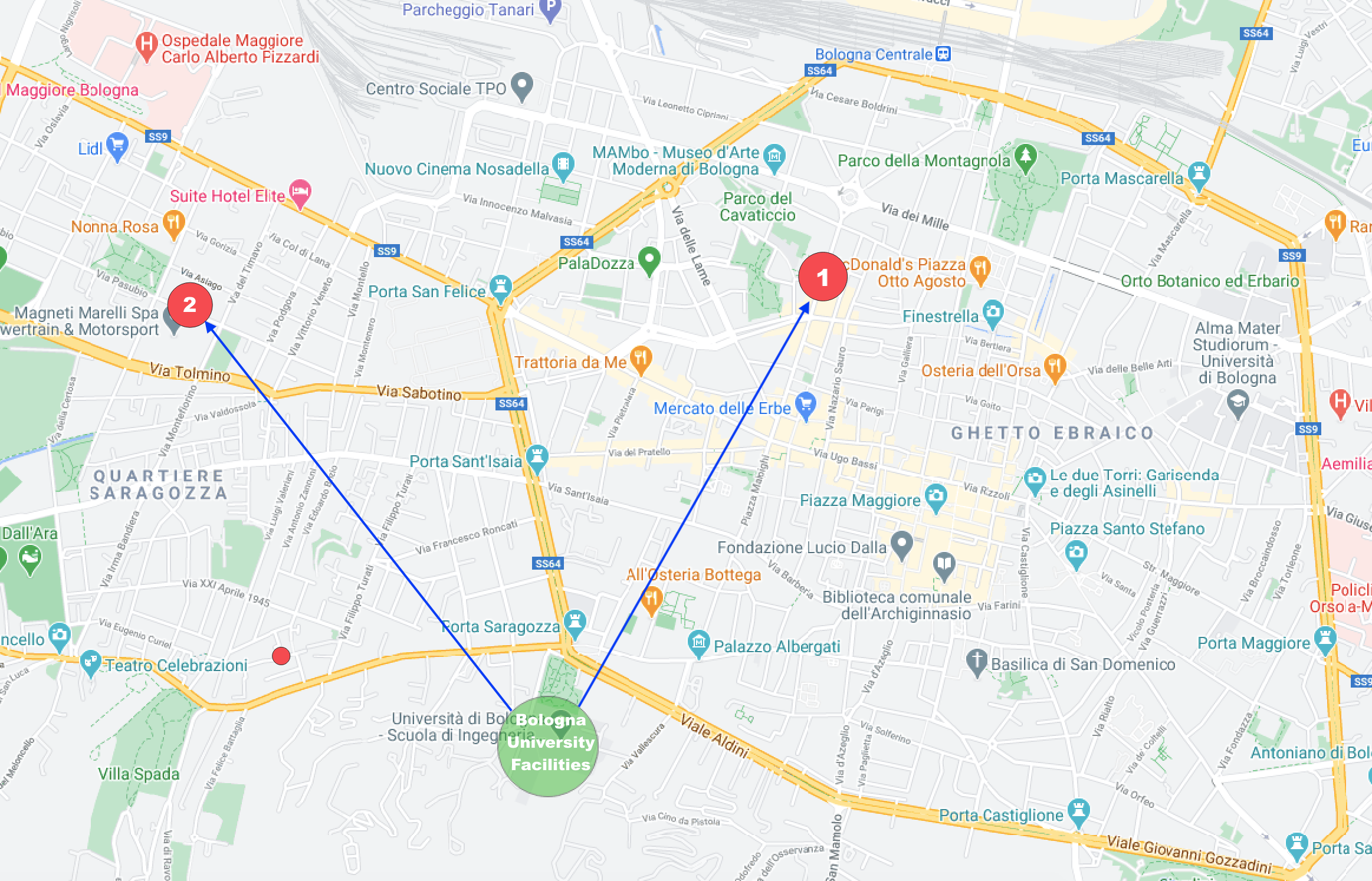}
    \label{fig:unibo_hub_b}
	}
\caption{Spreading out pattern from the engineering department hub, difference between May (a) and August (b).}
  \label{fig:unibo_hub}
\end{figure}

\item Similar to the findings of Zhao et al. \cite{zhao2015exploring}, we find out many people commute for work or studies and use bikes to arrive at the \textbf{train station} or to ride from the train station to their work or study place. In May, the four main destinations were various departments of the university (Figure \ref{fig:train_hub_a}), while in August (Figure \ref{fig:train_hub_b}), the final destinations changed and most of the trips ended in a large shopping mall just outside the city centre.

\item \noindent Considering the \textbf{engineering department}, it is interesting to note how the pattern changed from May to August. During the month of May, when the lessons are still in progress, a significant number of trips ended in places in the city where students spend their non-university time, such as student residences and pubs (Figure \ref{fig:unibo_hub_a}). Whereas, for the month of August, the two most popular destinations turned out to be near to parks (Figure \ref{fig:unibo_hub_b}). In addition, the number of trips also became very less in the month of August due to the holiday period.
\end{enumerate}

\subsection{Weather's effects on bike usage}\label{sec:weaAna}
In this section, we discuss the results about how weather conditions, that is air temperature, precipitations and wind speed, affect the number of bike trips during the six months period.

\subsubsection{\textbf{Air temperature}}
\begin{figure}[h]
    \centering
    \includegraphics[width=1\columnwidth]{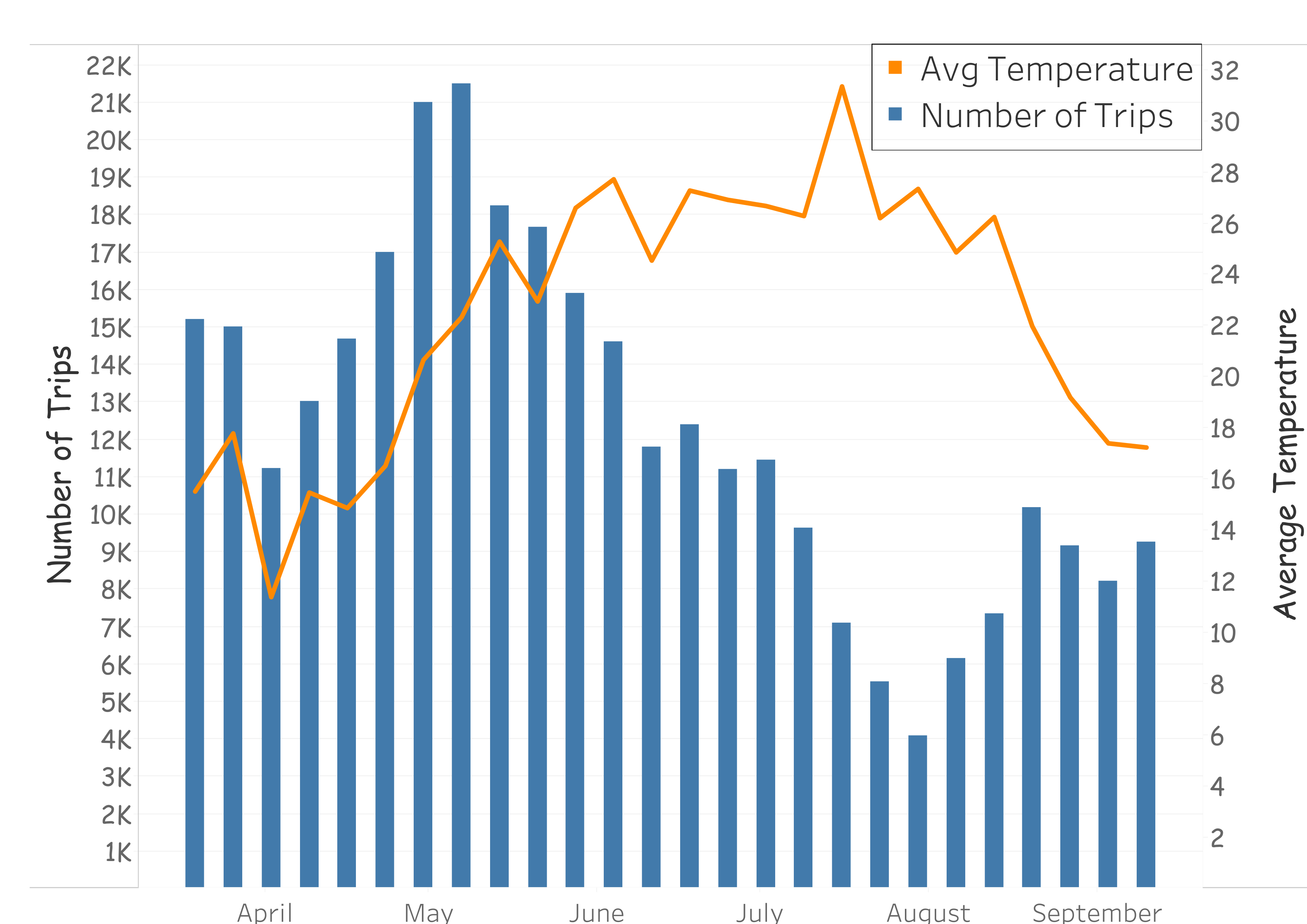}
    \caption{Average daily air temperatures and the number of bike trips.}
    \label{fig:Rides_temp2}
\end{figure}

The spring in Bologna is characterized by comfortable temperatures, while in the summer the weather is hot and sultry. Comfortable weather conditions usually play a significant role in bike usage, and when the air temperature is too high or too low, people prefer a different means of transport \cite{nosal2014effect}. Figure \ref{fig:Rides_temp2} shows the number of bike trips and the trend of the temperature during the six months period. Temperatures in April and May, which is around 20 degrees, is comfortable for making bike trips, while from mid of June to the end of August, when the average air temperature reaches 27 degrees or above, the number of trips starts to drop. In summary, we can confirm that bike trips are strongly dependant on air temperature, which confirms the findings of many previous studies \cite{nosal2014effect}, \cite{nankervis1999effect}.

\subsubsection{\textbf{Precipitations}}
\begin{figure}[h]
    \centering
    \includegraphics[width=1\columnwidth]{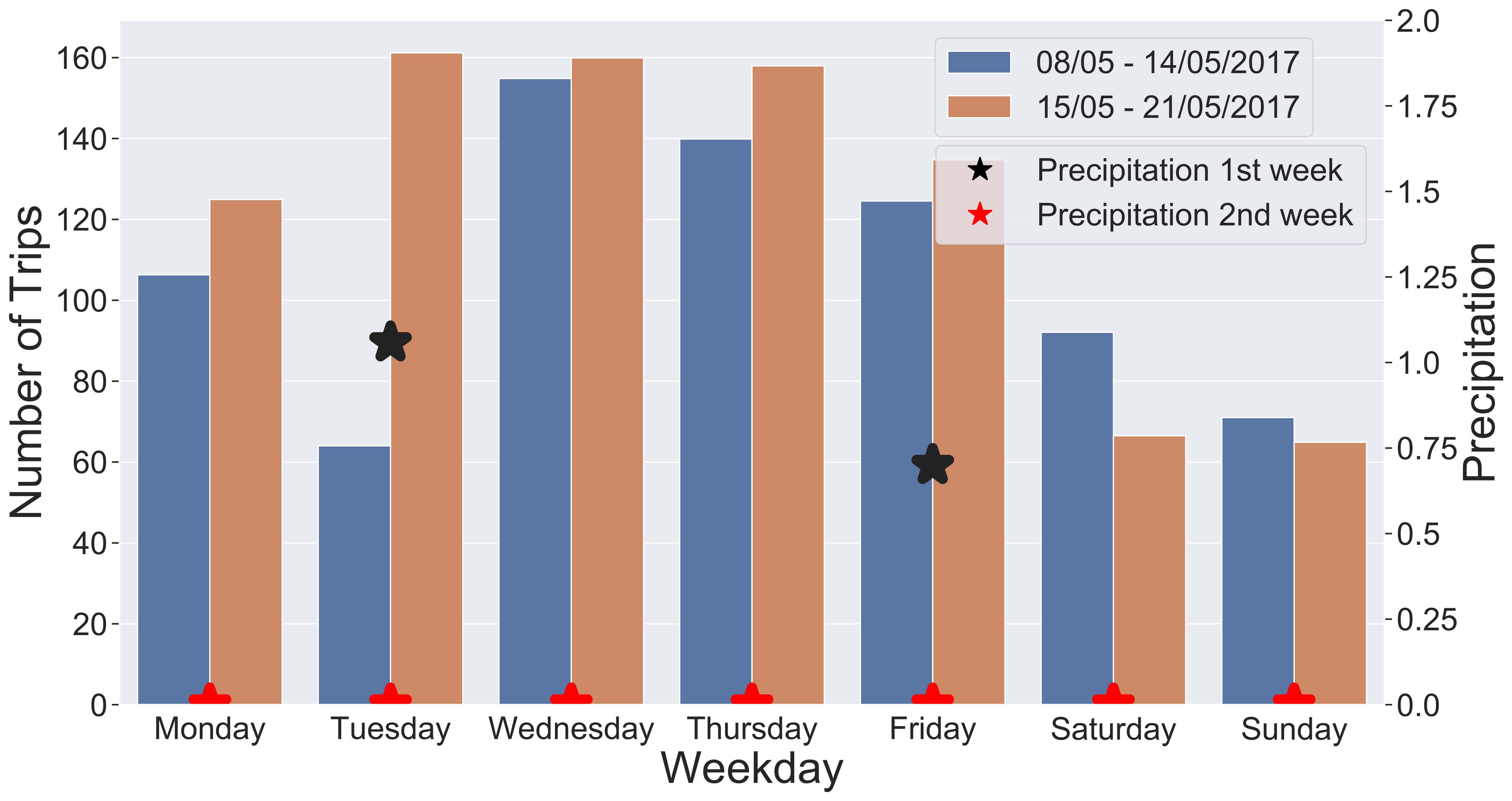}
    \caption{Precipitation amount and number of trips during different weeks in May.}
    \label{fig:May_comparison}
\end{figure}

\begin{figure}[h]
    \centering
    \includegraphics[width=1\columnwidth]{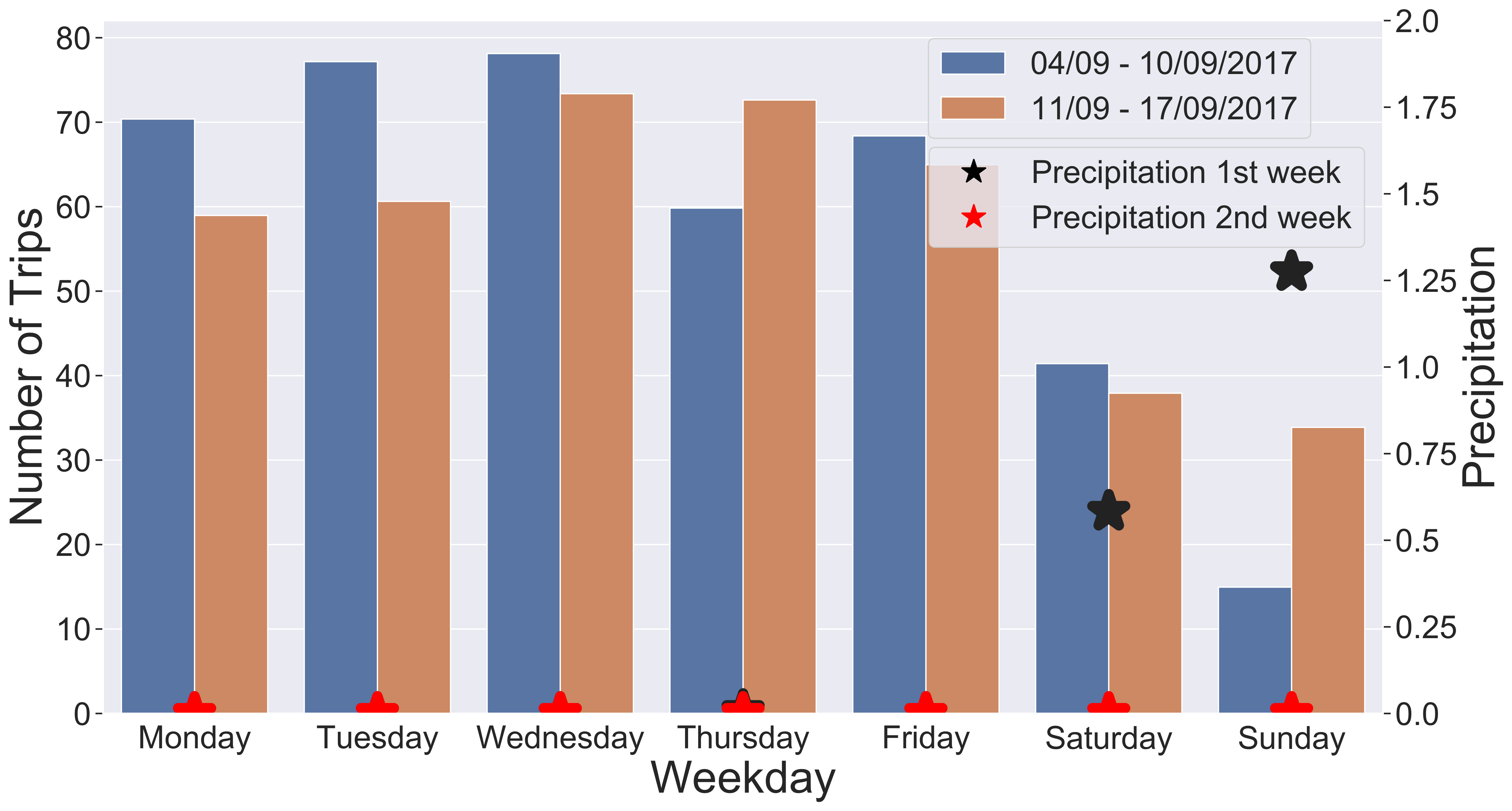}
    \caption{Precipitation amount and number of trips during different weeks in September.}
    \label{fig:Sept_comparison}
\end{figure}

During the six months of the observation period, most of the days were dry with no rain or snow. There were only some days from May 08 to May 14 and from September 04 to September 10 with some amount of precipitations, while during other months, the number of rainy days were significantly low. Thus, we decided to look into the weeks with the highest number of precipitations and compare them with the following week. In particular, we compared the week from May 08 to May 14 against the week from May 15 to May 21 (Figure \ref{fig:May_comparison}), and the week from September 04 to September 10 against the week September 11 to September 17 (Figure \ref{fig:Sept_comparison}).

From May 08 to May 14, there were two rainy days, that is Tuesday and Friday, and they heavily affected the number of bike trips (Figure \ref{fig:May_comparison}). In particular, on Tuesday the number of trips were less than half compared the next week's Tuesday. However, there was a small difference between the number of trips during Fridays. It is to be noted that most of the trips were made in the early morning hours, as shown in Figure \ref{fig:Daily trend}. Looking into hourly data, we detected that on May 09, there were heavy rains from 2 am to 9 am that strongly affected the number of trips, while on May 12, it was raining only for one hour around 3 pm, and this did not significantly affect the number of trips.

\noindent The same goes for the comparison between the two weeks of September (Figure \ref{fig:Sept_comparison}). In particular, from September 04 to September 10, there were two rainy days, that is Saturday and Sunday. On Saturday, the number of trips remained unchanged compared to the Saturday of the following week since it was raining from 8 pm to 10 pm, while on Sunday due to rain, which occurred from 2 am to 11 am, it led to a half number of trips.

\subsubsection{\textbf{Wind speed}}\label{sec:subsec_wind}
\begin{figure}[h]
    \centering
    \includegraphics[width=1\columnwidth]{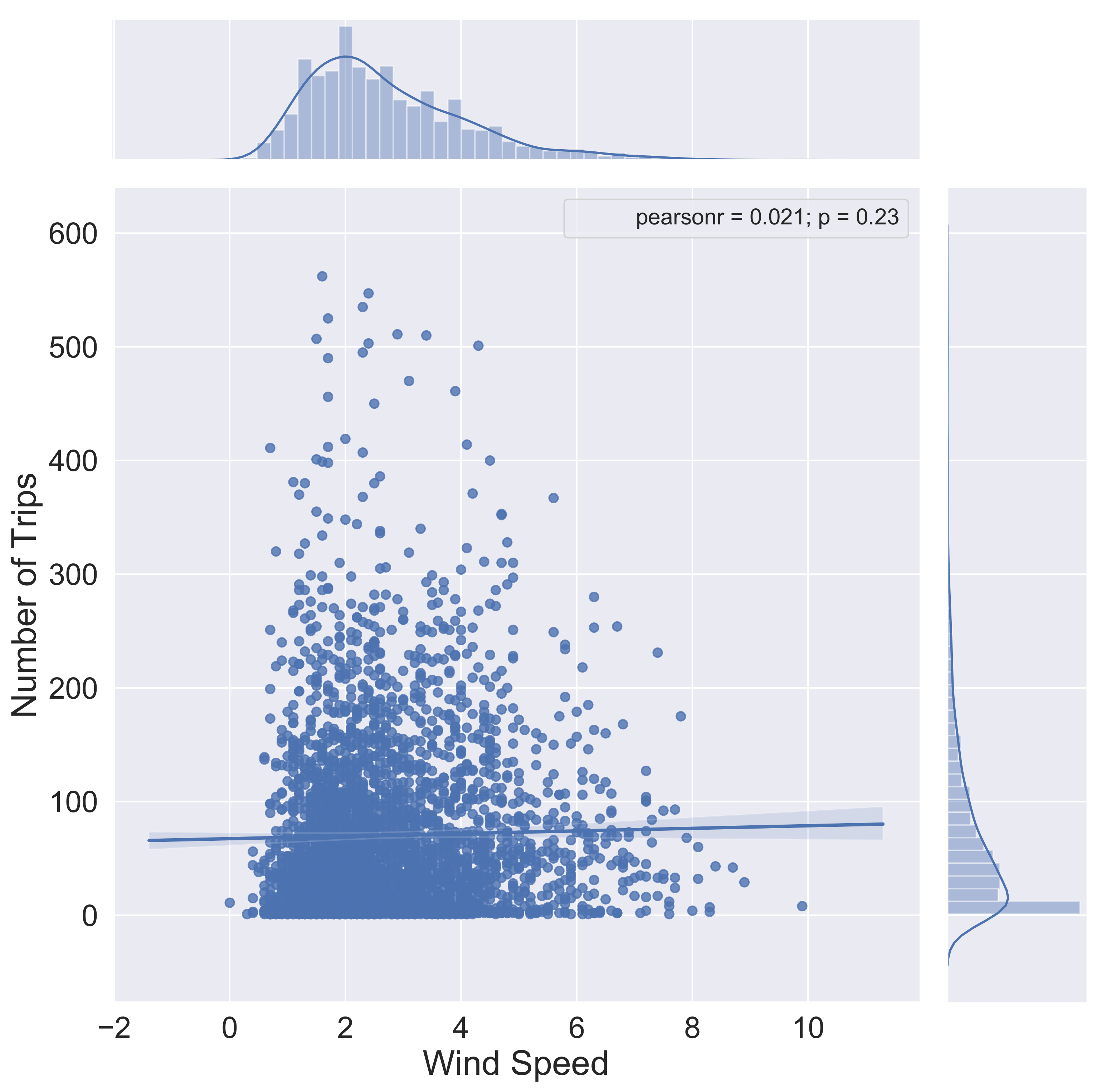}
    \caption{Average wind speed and the number of bike trips.}
    \label{fig:Wind}
\end{figure}

Finally, we analysed how the speed of wind affects the bike usage. Figure \ref{fig:Wind} shows the number of trips and the average wind speed during the six months period. The results show that there is no correlation between wind speed and number of bike trips during the observation period. 

\subsection{Pollution and bike usage}\label{sec:subsec_pollution}
Being one of the most ecological means of transportation, there is a notion that the usage of bikes might decrease the pollution of air \cite{hertel2008proper}, \cite{johansson2017impacts}. We analysed the changes in four different indicators of air pollution during the period of observation, that is i) Particulate matter (PM), ii) Ozone (O3), iii) Nitrogen dioxide (NO2), and iv) Sulfur dioxide (SO2). However, no positive or negative correlation could be identified. We also analysed if there were any changes in the indicators for separate streets that were being heavily used by bicycles. Similarly, we could not find any correlation between air pollution and bike use during the period of observation. Possibly reasons for this result could be due to having a low amount of observations about air pollution and the lack of car usage data.

\subsection{Holidays and events}\label{sec:eveAna}
\begin{figure}[t]
	\subfloat[]{
		\includegraphics[width=0.45\textwidth]{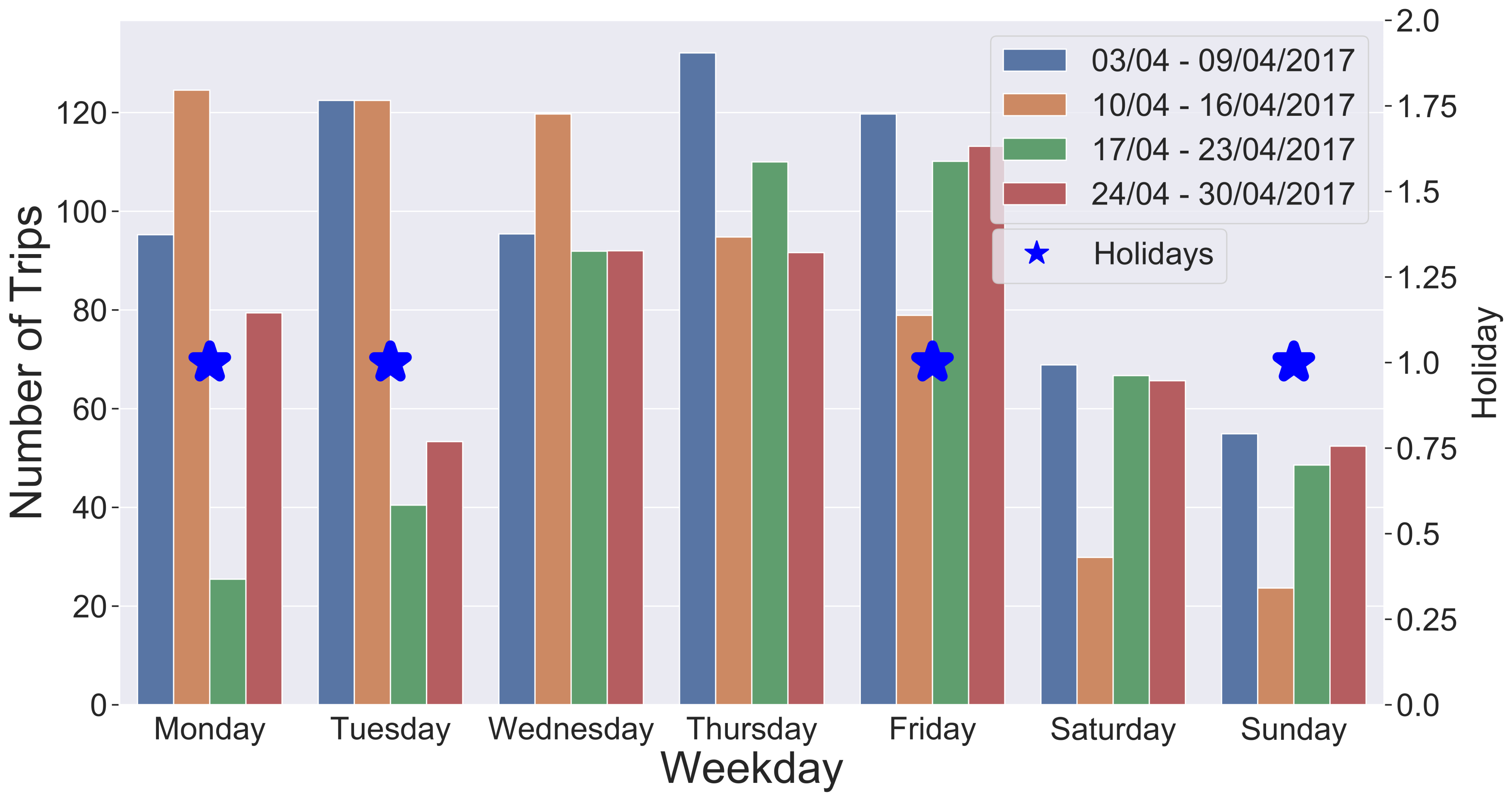}\label{fig:Hol_april}
	}~\hfill \\
	\subfloat[]{
	\includegraphics[width=0.45\textwidth]{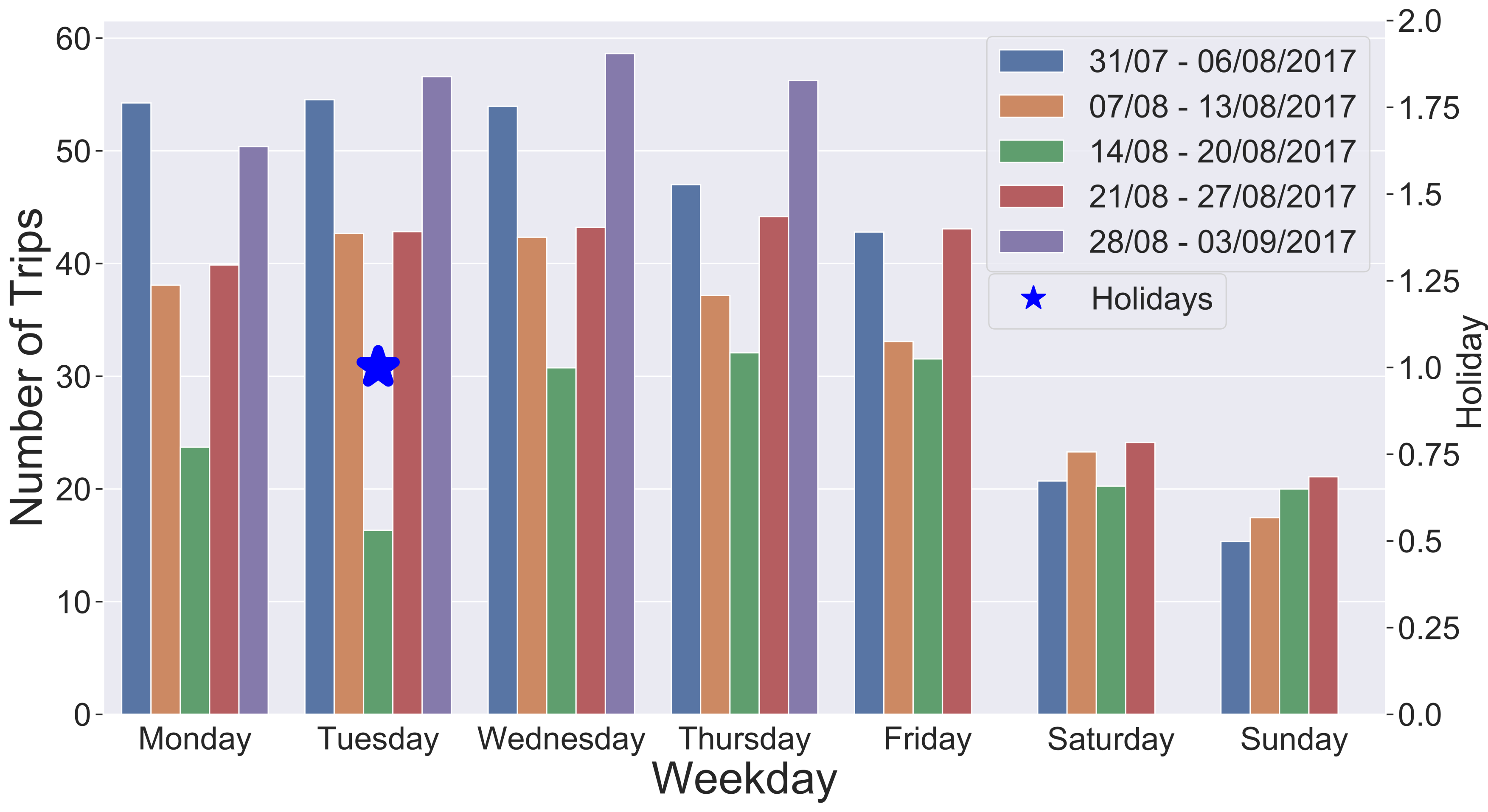}\label{fig:Hol_august}
	}
 \caption{Number of bike trips aggregated by day of the week for each week, for April (a) and August (b). Holidays are marked with a blue star.}\label{fig:Hol_rides}
\end{figure}

In this last section, we discuss how holidays, events, such as protests and strikes, and national celebration affect the bike usage. During the period of observation, there were 14 public holidays. Figure \ref{fig:Hol_rides} shows the bike trips aggregated by day of the week for April (Figure \ref{fig:Hol_april}) and September (Figure \ref{fig:Hol_august}), where the changes in the bike usage were more evident. In particular, the results show a significant difference during the Easter holidays (April 16-17, 2017) and during Ferragosto\footnote{Assumption of Mary day.} (i.e., August 15, 2017). Figure \ref{fig:Hol_april} shows that during Easter holidays, the number of trips dropped by 35\% on Good Friday, over by 50\% on Holy Saturday and Easter Sunday, that is the orange bars in the figure, and almost by 80\% on Easter Monday, that is the green bar in the figure. Figure \ref{fig:Hol_august} shows that during Ferragosto in August 2017, that is the green bar in the Figure, the number of trips dropped by 60\% in comparison to the number of trips done one week before and after.

Next, we analysed how events such as protests, strikes, and large gatherings in the city affect the bike usage. The reason behind analyzing the impact of these type of events is the following: during protests or strikes, the number of main roads are being diverted, blocked, or closed, which may lead to an increased number of bike trips that represents a more agile means of transportation. On the other hand, a strike may also decrease the usage of bikes as people may decide not to go to work. Analyzing the protests in Bologna during the six months period, we did not notice any significant change in the usage of the bike. It is worth mentioning that most of the protests and strikes were not large and did not last for more than a few hours. Also, although no information was found on the closure of the roads, the number of bike trips did not significantly change during these events.

%% file: 05.predictions.tex
In this section, we present the results of the predictive analysis with the aim to predict bike trips for the next 30 and 60-minutes. 

\subsection{Predictive algorithms and Metrics: } We employed various methods, that is Linear Regression, Random Forest, Extreme Gradient Boosting (XGBoost), and LSTM for the predictive task. The prediction models are evaluated using different metrics that allow understanding the performance of the predictive algorithms. In particular, given the regression nature of the prediction, we used the following metrics: Mean Absolute Error (MAE), Mean Squared Error (MSE), Root Mean Squared Error (RMSE), and R$^2$ metrics.

\begin{figure*}[t!]
    \centering
    \includegraphics[width=0.75\textwidth]{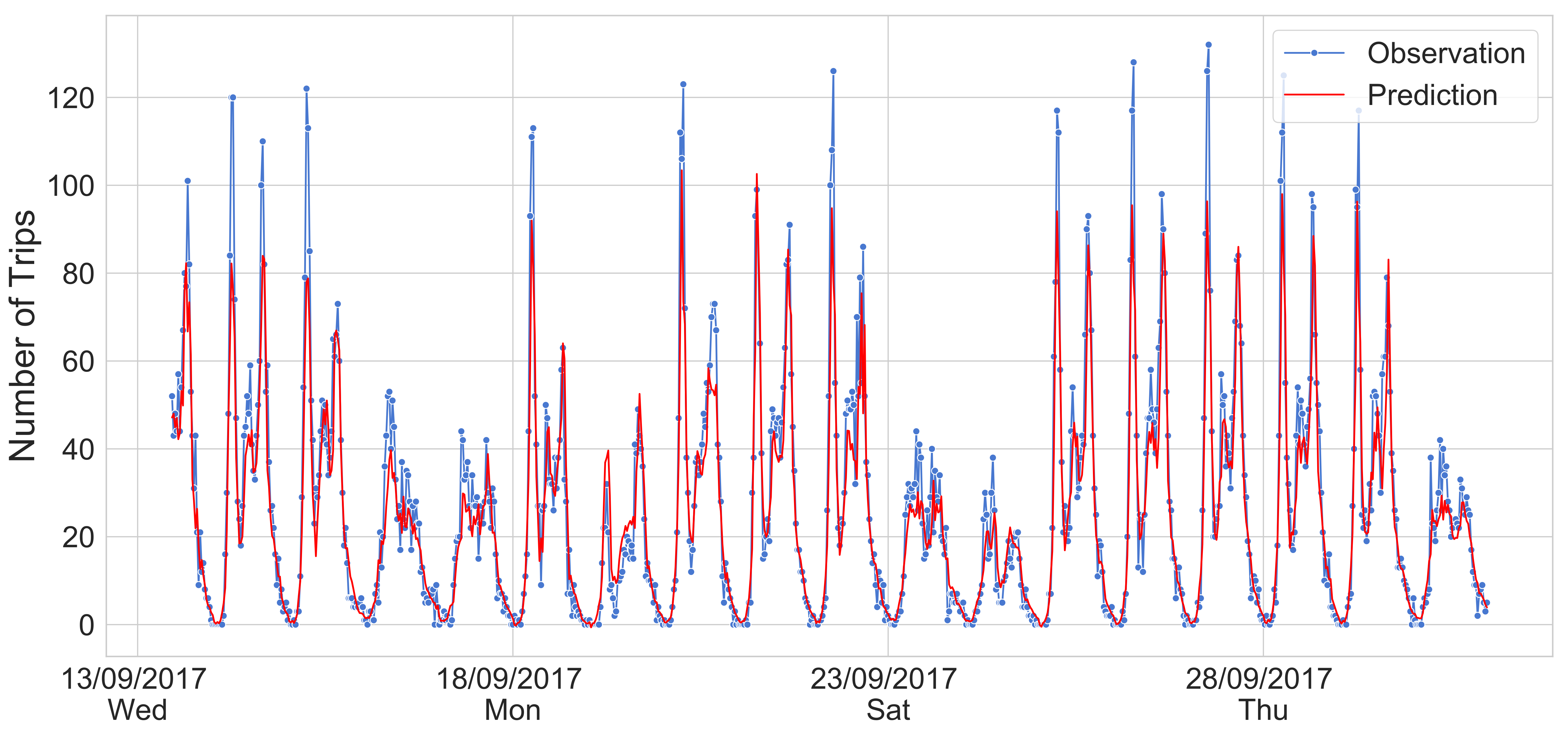}
    \caption{LSTM model 30-minute prediction over the test set of 18-day (September 13, 2017 - September 30, 2017).}
    \label{fig:LSTMprediction}
\end{figure*}

\subsection{Feature selection}\label{features}
We manually selected the features for the prediction. In particular, we selected the following features: \emph{temperature}, \emph{precipitation}, \emph{hour\_of\_the\_day}, \emph{month}, \emph{season} (spring, summer and autumn), \emph{day\_of\_week} (Monday, Tuesday, ..., Sunday), \emph{holiday} (whether a day is a holiday or not), \emph{hour\_history}, and \emph{week\_history}. The last two variables represent the number of trips for one hour and one week before, respectively. Based on the results shown in Sections \ref{sec:subsec_wind} and \ref{sec:subsec_pollution}, we excluded wind speed and air pollution in our final evaluation.

\subsection{Experimental setup}\label{Experiment Setup}
In order to predict bike usage in the next 30 and 60-minutes, we have prepared two datasets. In the first one,  we aggregated the number of trips by 30-minutes time slots, while in the second one, we aggregated the number of trips by hours. In both datasets, we converted categorical variables into dummy variables. We computed the correlation among all the variables with the number of trips, and we found out that the number of trips is mostly correlated with \emph{hour\_history}, \emph{week\_history}, \emph{temperature}, \emph{month}, and \emph{season} (i.e., spring, summer, and autumn). In addition, we used four different splits that is, 90/10, 80/20, 70/30, and 60/40 for training/test ratio. For each split, we applied the 10-fold Cross-Validation technique on the training set and evaluate it on the test data.

\subsection{Prediction results}\label{Prediction Results}
Among all the prediction models, LSTM gave the best results\footnote{We are not showing all the results due to space limitation.}. Table \ref{table:models_comparison} provides the results for the next 60-minutes interval prediction for the 90/10 split. We can notice that the LSTM model is almost 4.5 times better than the others, and the second-best model is the Random Forest, which is slightly better than Linear Regression and XGBoost. Thus, considering the best performance of the LSTM model, we provide results with various data split ratios in Table \ref{table:lstm_splits}, both for the 60-minutes as well as for the 30-minutes interval prediction. The best results were achieved with 90/10 split ratio. It is worth to notice that the results improve significantly by reducing the prediction interval (30-minutes). In particular, predicting 30-minutes interval improves the results on average 4 times in comparison with 60-minutes predictions.
The importance of \emph{hour\_history} feature can be noted by that it improved the model on an average by 30\%, considering MSE, MAE and RMSE results. Whereas the \emph{week\_history} feature improved the results by around 45\%. As a micro analysis, Figure \ref{fig:LSTMprediction} shows the LSTM prediction results over the 18 days of the test set (September 13, 2017 - September 30, 2017). Each time step on the graph is 30 minutes and it is worth noticing that the prediction trend (i.e., the red line in the graph) almost completely overlaps the observation trend (i.e., the blue line in the graph).

\begin{table}[h!]
\centering
  \begin{tabular}{|r | c c c c|}
    \hline
    \textbf{Method} & \textbf{MAE} & \textbf{MSE} & \textbf{RMSE} & \textbf{R$^2$} \\
    \hline
    Linear Regression & 46.03 & 4438.29 & 66.62 & 0.25\\
    Random Forest & 44.72 & 4354.25 & 65.98 & 0.33\\
    XGBoost & 47.04 & 4809.13 & 69.34 & 0.17\\
    LSTM & \textbf{16.62} & \textbf{457.73} & \textbf{21.39} & \textbf{0.91}\\
    \hline
  \end{tabular}
  \caption{Models comparison in 60-minutes interval prediction using 90/10 split ratio.}
  \label{table:models_comparison}
\end{table}

\begin{table}[h!]
  \resizebox{\columnwidth}{!}{
  \begin{tabular}{| c | c | c | c | c | c | c | c | c |}
    \hline
    \textbf{Split} & \multicolumn{2}{c|}{\textbf{MAE}} & \multicolumn{2}{c|}{\textbf{MSE}} & \multicolumn{2}{c|}{\textbf{RMSE}} & \multicolumn{2}{c|}{\textbf{R$^2$}} \\
    \cline{2-9}
    \textbf{ratio} & \emph{60} & \emph{30} & \emph{60} & \emph{30} & \emph{60} & \emph{30} & \emph{60} & \emph{30}\\
    \hline
    90/10 & 16.62 & 5.38 & 457.73 & 66.08 & 21.39 & 8.12 & 0.91 & 0.91\\
    80/20 & 23.62 & 6.42 & 986.56 & 104.54 & 31.40 & 10.22 & 0.63 & 0.86\\
    70/30 & 25.44 & 5.62 & 1140.87 & 70.63 & 32.45 & 8.40 & 0.70 & 0.87\\
    60/40 & 16.75 & 6.03 & 546.67 & 98.64 & 23.38 & 9.13 & 0.75 & 0.85\\
    \hline
  \end{tabular}
  }
  \caption{Prediction results varying the split ratio. The values represents the LSTM model results predicting 60-minutes/30-minutes interval, respectively.}
  \label{table:lstm_splits}
\end{table}

%% file: 06.conclusion.tex
The significantly growing awareness about climate change and pollution has given rise to the need for eco-friendly and healthy means of transportation. Cycling allows lightening road traffic within touristic and historical cities where the traffic congestion is exacerbated. Thus, understanding cycling mobility is of the utmost importance to improve bike infrastructures and encourage bike use.

In this paper, we presented a descriptive and predictive analysis of bike usage for the city of Bologna. In particular, we analysed temporal and spatial patterns of bike users and the impact of weather conditions, such as air temperature, precipitation, and wind speed, pollution conditions, and holidays on the bike usage. Moreover, we compared different models to predict the trips for the short-term period, that is, 30-minutes and 60-minutes time interval. The results show a seasonality of cycling trips and more in detail a weekly trend in favour of working days, which is congruent with a commuting behaviour from/to work and study place. Also, the spatial analysis confirms this result, in particular, we found several attractive points that coincide with places of study, which were less frequently visited during the summer period. The several supplementary datasets used in the descriptive analysis allowed to confirm a negative correlation between bike trips and precipitation and highlight that temperature around 26/27 degrees and over leads to a decrease in the number of bike trips. However, we did not find evidence about the correlation between air pollution and bike usage. In the predictive analysis, we found that LSTM provides the best results and that too for predicting for 30-minutes (compared to 60-minutes) time interval, which could be of practical help to management of traffic infrastructure (e.g., traffic lights, temporary traffic detours) on road network links. Additionally, the model can be used to predict the demand gap for strengthening the bike-sharing services, which usually require a redistributing service to make sure that people who would like to use bikes will most certainly find one near their location.

As future works, several directions can be considered. Firstly, we plan to analyse the other transport data within the \emph{Bella Mossa} dataset to understand interactions with bike usage and different behaviour in the use of the city's road network. We also plan to study a larger dataset with a longer timeline to improve analysis of seasonal factors that may also improve the prediction results. Finally, we plan to compare the patterns of bike users from different cities.